\documentclass{article}
\usepackage[all]{xy}
\usepackage{amsfonts}
\usepackage{amsmath}
\usepackage{eucal}
\usepackage{amssymb}

\newcommand{\eq}{\begin{equation}}
\newcommand{\feq}{\end{equation}}
\newcommand{\eqn}{\begin{eqnarray}}
\newcommand{\feqn}{\end{eqnarray}}
\newcommand{\arr}{\begin{eqnarray*}}
\newcommand{\farr}{\end{eqnarray*}}

\newtheorem{prop}{Proposition}

\begin{document}

\begin{titlepage}
\begin{flushright}
IFUM--827--FT\\
LBNL--57265\\
UCB--PTH--05/05
\end{flushright}
\vskip 10mm
\begin{center}
\renewcommand{\thefootnote}{\fnsymbol{footnote}}
{\Large \bf Euler angles for \boldmath{$G_2$}}
\vskip 12mm
{\large \bf {Sergio L.~Cacciatori$^{1,3}$\footnote{cacciatori@mi.infn.it},
Bianca L. Cerchiai$^{4,5}$\footnote{BLCerchiai@lbl.gov},
Alberto Della Vedova$^{2}$\footnote{alberto.dellavedova@unimib.it}
Giovanni Ortenzi$^{2}$\footnote{giovanni.ortenzi@unimib.it},
Antonio Scotti$^1$\footnote{Graduate visitor (visitatore laureato), 
ascotti@mindspring.com}}}\\
\renewcommand{\thefootnote}{\arabic{footnote}}
\setcounter{footnote}{0}
\vskip 5mm
{\small
$^1$ Dipartimento di Matematica dell'Universit\`a di Milano,\\
Via Saldini 50, I-20133 Milano, Italy. \\

\vspace*{0.2cm}

$^2$ Dipartimento di Matematica ed Applicazioni, Universit\`a
Milano Bicocca,\\
via  R. Cozzi, 53- I-20126 Milano, Italy. \\



\vspace*{0.4cm}

$^3$ INFN, Sezione di Milano,\\
Via Celoria 16,
I-20133 Milano, Italy.\\

\vspace*{0.4cm}

$^4$ Lawrence Berkeley National Laboratory\\
Theory Group, Bldg 50A5104\\
1 Cyclotron Rd, Berkeley, CA 94720-8162, USA.\\

\vspace*{0.4cm}

$^5$ Department of Physics, University of California,\\
Berkeley, CA 94720, USA.
}
\end{center}
\vspace{1cm}
\begin{center}
{\bf Abstract}
\end{center}
{\small  We provide a simple coordinatization for the group $G_2$, which
is analogous to the Euler coordinatization for $SU(2)$. We show how 
to obtain the general element of the group in a form emphasizing 
the structure of the fibration of $G_2$ with fiber $SO(4)$ and base
$\mathcal{H}$, the variety of quaternionic subalgebras of octonions.  
In particular this allows us to obtain a simple expression for the Haar 
measure on $G_2$. Moreover, as a by-product it yields a concrete 
realization and an Einstein metric for $\mathcal{H}$.} 

\end{titlepage}


\section{Introduction}

The relevance of Lie groups in physics is a well established fact: They
appear both in classical and in quantum problems. In this context an important 
role is played by the Haar measure, needed e.g. for the 
construction of a consistent path integral in lattice gauge theories 
\cite{Rothe}. The canonical 1-form $\theta$ of 
the compact Lie group $G$ is the fundamental structure customarily used to find this measure. 
In effect the components  of $\theta$ w.r.t. a basis of $Lie(G)$ are 
everywhere linear independent, smooth, left-invariant 1-form on $G$ \cite{Simon}. So their wedge product
gives us a left-invariant volume form. 
In order to perform explicit calculations we have to choose a suitable 
local chart and the related local expression for the Haar measure. 
The logarithmic coordinates \cite{DK} are the most obvious choice for 
a coordinatization of $G$. In this case the canonical 1-form becomes 
$$      
\theta(X)=\int_0^1 e^{s \, \mathrm{ad}X} \ ds.
$$
The related volume form is
$$
\omega=\prod_{\lambda \in \sigma(\mathrm{ad}X)} \frac{1-e^{-\lambda}}{\lambda} d \alpha_1 \wedge 
\dots \wedge d \alpha_n,
$$
where $\sigma(\mathrm{ad}X)$ denotes the spectrum of $\mathrm{ad}X$ and $\alpha_i$ are a basis of
$Lie(G)^*$. However these coordinates do not display generally the subgroup structure of $G$ which 
usually are relevant in the physical applications. The difficulties of  such a kind of coordinatization
arise when one needs to explicitly determine the global range of the coordinates.
A coordinatization which yields a simple form for the Haar measure 
and at the same time allows a simple determination of the range for the 
angles can become crucial for numerical computations, 
e.g. in lattice gauge theories or in random matrix models.
For unitary groups such a coordinatization has been constructed in \cite{Tilma},
generalizing the ``Euler angle parameterization'' for $SU(2)$.

In this paper we provide an analogous simple coordinatization for the 
exceptional Lie group $G_2$. We start by showing how a simple matrix 
realization of the algebra can be obtained starting from the octonions. 
Then a proposal for a representation of the group elements, based 
on the previous construction and emphasizing the $SO(4)$ subgroup 
embedded in $G_2$, is given and it is proven to cover the whole group.

After computing the left-invariant currents in a way that respects the 
structure of the fibration, the infinitesimal invariant measure is determined 
with a suitable normalization. We then use
topological tools and symmetry arguments to determine the correct range 
of the coordinates. As a by-product we obtain a coordinatization and an 
Einstein metric for the eight-dimensional variety of quaternionic
subalgebras of octonions.

Motivations for considering models with $G_2$ symmetries are provided 
by different physical systems, for example they arise in the study of
deconfinement phase transitions \cite{Pepe}, in random matrix models 
\cite{Zeev} or in the new matrix models related to $D-$brane 
physics \cite{Krauth:2000bv}.


\section{The \boldmath{$G_2$} algebra}
\setcounter{equation}{0}

The octonions $\mathbb{O}$ are an eight-dimensional real algebra whose
generic element $a$ is a pair of quaternions $(\alpha_1, \alpha_2)$ 
with the following multiplication rules
\eqn
\label{moltott}
 (\alpha_1,\alpha_2)\cdot(\beta_1,\beta_2)&=& (\alpha_1
 \beta_1-\bar \beta_2 \alpha_2, \beta_2 \alpha_1 + \alpha_2 \bar
 \beta_1).  
\feqn
Here $(\alpha_1,\alpha_2),(\beta_1,\beta_2)$ are generic octonions.
This algebra comes naturally equipped with an involution called conjugation 
$$
\overline{(\alpha_1,\alpha_2)}=(\bar \alpha_1,-\alpha_2).
$$
Denoting by $1,i,j,k$ the usual basis of the quaternions yields the
following canonical basis for the algebra $\mathbb{O}$
$$ e_0=(1,0), \:\: e_1=(i,0), \:\: e_2=(j,0), \:\: e_3=(k,0), 
\:\: e_4=(0,1), \:\: e_5=(0,i), \:\: e_6=(0,j), \:\: e_7=(0,k). $$
Using this basis it follows easily that the subspace $\mathbb H$ 
spanned by $\{ 1,e_1,e_2,e_3 \}$ is in fact a quaternionic 
subalgebra that we call the canonical quaternionic subalgebra.
Moreover we can consider $\mathbb{O}$ as a two-dimensional module over 
$\mathbb{H}$, i.e. every octonion $z$ can be decomposed as $z=x+ye_4$,
where $x,y$ are suitable quaternions.\\  
The octonions, together with $\mathbb{R},\mathbb{C}$ and $\mathbb{H}$, are 
the only normed division algebras. The norm is induced by the 
standard Euclidean structure of the underlying real vector space. They are 
neither commutative nor associative, but they are 
alternative, i.e. any subalgebra generated by two octonions is associative. 
This weak form of associativity implies the Moufang identities \cite{schafer}, 
which are multiplication laws among octonions and will prove very useful 
in the following
\eqn
\label{Moufang}
(ax)(ya)&=&a(xy)a, \cr
a(x(ay))&=&(axa)y, \cr
y(a(xa))&=&y(axa).
\feqn 
The relevance of the octonions in mathematics is due to their deep connection 
with the exceptional Lie groups (\cite{baez} and references therein). 
We are interested in the group $G_2$. In this case the link is easy to 
understand: $G_2$ is the automorphism group of the octonions. 
For every octonion $a$ we denote by $l_a$ the left multiplication $l_a(x)=ax$.
Under suitable hypothesis the composition of left multiplications generates 
elements of $G_2$.
In fact it holds 
\begin{prop}
\label{carg2}
Let $g=l_{a_1} \dots l_{a_n}$, where $a_1,\ldots,a_n$ are unitary 
purely imaginary octonions. If $g(1)=1$, then $g \in G_2$.   
\end{prop}

\textit{Proof }
We have to show that $g(x)g(y)=g(xy)$ for all $x,y \in \mathbb{O}$. To
this end we prove by induction on $n$ that 
\begin{equation}\label{spin7}
  g(x)g(y)=g((xb)y), 
\end{equation} where
$b=(\ldots(a_1a_2)\ldots)a_n$. 

For the moment we avoid the hypothesis $g(1)=1$.

If $n=1$ we have $$(ax)(ay)=-a^2((ax)(ay))=-a((a(ax)a)y)=a((xa)y),$$
where the first equality holds because $a$ is purely imaginary and the
others by the Moufang identities.

Now we suppose the statement is true for $n$. So we have 
$$g(ax) g(ay) = g(((ax)b)(ay)) = -g(a^2(((ax)b)(ay))) = g(a((x(ba))y)),$$ 
which is the equation (\ref{spin7}), if we replace $g$ by $g\, l_a$ and $b$ 
by $ba$.

Finally, to complete the proof we have to show that
$b=(\ldots(a_1a_2)\ldots)a_n=1$. 
So, applying the operator $l_{a_n}\ldots l_{a_1}$ to both members of
the equation $1=g(1)=a_1(\ldots(a_{n-1}a_n)\ldots)$ gives us 
$ a_n(a_{n-1}(\ldots(a_2a_1)\ldots))=(-1)^n $, and then by conjugation we 
obtain
$(-1)^n (\ldots(a_1a_2)\ldots)a_n=(-1)^n$ which implies $b=1$.
\begin{flushright} $ \square$ \end{flushright}

We can get some interesting subgroups of $G_2$ by imposing additional 
conditions on $g$. If we add the hypothesis $g(e_1)=e_1$ we obtain a 
$SU(3)$ subgroup denoted in the following by $P$. 
Moreover, imposing also $g(e_2)=e_2$, the resulting subgroup is a copy of 
$SU(2)$, which we call $S$.

In order to write down generalized Eulerian coordinates we need to 
describe the embedding of $SO(4)$ in $G_2$ \cite{conway}.

To this end we identify as usual $SU(2)$  with the $3$--sphere of unitary 
quaternions and we consider the following homomorphism
\eqn
\label{omomo}
\gamma: SU(2) \times SU(2) &\to& G_2 \cr
(a,b)&\mapsto& \gamma_{ab},
\feqn
where $\gamma_{ab}(x+ye_4)=a x \bar{a} + (b y \bar{a}) e_4$. 
Using the Moufang identities it is not hard to check that $\gamma_{ab}$ 
is truly an octonion automorphism.
Fixing $a=1$ provides us with the embedding $\gamma_{1b}$ of $SU(2)$ in $G_2$, 
whose image is the subgroup $S$.
On the other hand the image of the embedding $\gamma_{a1}$ is a $SU(2)$, 
which we denote by $\Sigma$ and which is not conjugate to $S$.

The map $\gamma$ is not injective and its kernel is the subgroup 
$\mathbb{Z}_2= \{(1,1),(-1,-1) \}$.
By the homomorphism theorem the image of the map $\gamma$ in $G_2$ is 
isomorphic to $(SU(2)\times SU(2))/\mathbb{Z}_2$, which is $SO(4)$ 
as well known.
In the following sections we will refer to the image of $\gamma$ simply as 
$SO(4)$.

The homogeneous space $\mathcal H = G_2 / SO(4)$ is the eight-dimensional 
variety of the quaternionic subalgebras of $\mathbb{O}$. So $\mathbb H$ 
can be thought as a point of $\mathcal H$. $G_2$ acts 
transitively on $\mathcal H$ and the stabilizer of
$\mathbb H$ is the image of $\gamma$ \cite{conway}. Then we have the fibration 
$$
\xymatrix{SO(4)  \ar@{^{(}->}[r] & G_2 \ar[d]\\
 & \mathcal{H}}
$$

The $G_2$ generators provided in proposition \ref{carg2}
are useful in particular to find a basis of the Lie algebra $Lie(G_2)$. 
Actually, let us take an element of $G_2$ of the form 
$g_{abc}=-l_{(cb)a}l_a l_b l_c$, where 
$(cb)a,a,b,c$ are unitary and purely imaginary (i.e. $a+\bar
a=0$). Notice that the choice of $-(cb)a$ guarantees  $g_{abc}1=1$. 
The condition that $(cb)a$ be purely imaginary amounts to $(cb) \perp a$.

Consider now a path $g_{a_tbc}=-l_{(cb)a_t}l_{a_t} l_b l_c$   
where $a_t=c \cos(t)+a\sin(t)$ and with  the additional requirements 
$b \perp c$, $b \perp a$ and $a \perp c$.

By definition $C_{abc}=\frac{d}{dt}\Big{\vert}_{t=0}g_{abc}(t)$ is an element 
of $Lie(G_2)$. 
With suitable choices of the elements $a,b$ and $c$ among the elements of the
canonical basis of $\mathbb O$ we can find a basis of this algebra. 
The representative matrices, written below with respect to the canonical basis,
are normalized with the condition $Tr (C_I C_J ) =-4 \delta_{IJ}$.  
We remark that they are seven-dimensional because of the trivial action on 
the real unity.
$$
C_1 =\left(
\begin{array}{ccccccc}
0 & 0 & 0 & 0 & 0 & 0 & 0 \\
0 & 0 & 0 & 0 & 0 & 0 & 0 \\
0 & 0 & 0 & 0 & 0 & 0 & 0 \\
0 & 0 & 0 & 0 & 0 & 0 & -1 \\
0 & 0 & 0 & 0 & 0 & -1 & 0 \\
0 & 0 & 0 & 0 & 1 & 0 & 0 \\
0 & 0 & 0 & 1 & 0 & 0 & 0 
\end{array}
\right)
\qquad
 C_2 =\left(
\begin{array}{ccccccc}
0 & 0 & 0 & 0 & 0 & 0 & 0 \\
0 & 0 & 0 & 0 & 0 & 0 & 0 \\
0 & 0 & 0 & 0 & 0 & 0 & 0 \\
0 & 0 & 0 & 0 & 0 & 1 & 0 \\
0 & 0 & 0 & 0 & 0 & 0 & -1 \\
0 & 0 & 0 & -1 & 0 & 0 & 0 \\
0 & 0 & 0 & 0 & 1 & 0 & 0 
\end{array}
\right)
$$
$$
C_3 =\left(
\begin{array}{ccccccc}
0 & 0 & 0 & 0 & 0 & 0 & 0 \\
0 & 0 & 0 & 0 & 0 & 0 & 0 \\
0 & 0 & 0 & 0 & 0 & 0 & 0 \\
0 & 0 & 0 & 0 & -1 & 0 & 0 \\
0 & 0 & 0 & 1 & 0 & 0 & 0 \\
0 & 0 & 0 & 0 & 0 & 0 & -1 \\
0 & 0 & 0 & 0 & 0 & 1 & 0 
\end{array}
\right)
\qquad
C_4 =\left(
\begin{array}{ccccccc}
0 & 0 & 0 & 0 & 0 & 0 & 0 \\
0 & 0 & 0 & 0 & 0 & 0 & 1 \\
0 & 0 & 0 & 0 & 0 & 1 & 0 \\
0 & 0 & 0 & 0 & 0 & 0 & 0 \\
0 & 0 & 0 & 0 & 0 & 0 & 0 \\
0 & 0 & -1 & 0 & 0 & 0 & 0 \\
0 & -1 & 0 & 0 & 0 & 0 & 0 
\end{array}
\right)
$$
$$
C_5 =\left(
\begin{array}{ccccccc}
0 & 0 & 0 & 0 & 0 & 0 & 0 \\
0 & 0 & 0 & 0 & 0 & -1 & 0 \\
0 & 0 & 0 & 0 & 0 & 0 & 1 \\
0 & 0 & 0 & 0 & 0 & 0 & 0 \\
0 & 0 & 0 & 0 & 0 & 0 & 0 \\
0 & 1 & 0 & 0 & 0 & 0 & 0 \\
0 & 0 & -1 & 0 & 0 & 0 & 0 
\end{array}
\right)
\qquad
C_6 =\left(
\begin{array}{ccccccc}
0 & 0 & 0 & 0 & 0 & 0 & 0 \\
0 & 0 & 0 & 0 & 1 & 0 & 0 \\
0 & 0 & 0 & -1 & 0 & 0 & 0 \\
0 & 0 & 1 & 0 & 0 & 0 & 0 \\
0 & -1 & 0 & 0 & 0 & 0 & 0 \\
0 & 0 & 0 & 0 & 0 & 0 & 0 \\
0 & 0 & 0 & 0 & 0 & 0 & 0 
\end{array}
\right)
$$
$$
C_7 =\left(
\begin{array}{ccccccc}
0 & 0 & 0 & 0 & 0 & 0 & 0 \\
0 & 0 & 0 & -1 & 0 & 0 & 0 \\
0 & 0 & 0 & 0 & -1 & 0 & 0 \\
0 & 1 & 0 & 0 & 0 & 0 & 0 \\
0 & 0 & 1 & 0 & 0 & 0 & 0 \\
0 & 0 & 0 & 0 & 0 & 0 & 0 \\
0 & 0 & 0 & 0 & 0 & 0 & 0 
\end{array}
\right)
\qquad
C_8 =\frac 1{\sqrt 3} \left(
\begin{array}{ccccccc}
0 & 0 & 0 & 0 & 0 & 0 & 0 \\
0 & 0 & -2 & 0 & 0 & 0 & 0 \\
0 & 2 & 0 & 0 & 0 & 0 & 0 \\
0 & 0 & 0 & 0 & 1 & 0 & 0 \\
0 & 0 & 0 & -1 & 0 & 0 & 0 \\
0 & 0 & 0 & 0 & 0 & 0 & -1 \\
0 & 0 & 0 & 0 & 0 & 1 & 0 
\end{array}
\right)
$$
$$
C_9 =\frac 1{\sqrt 3} \left(
\begin{array}{ccccccc}
0 & -2 & 0 & 0 & 0 & 0 & 0 \\
2 & 0 & 0 & 0 & 0 & 0 & 0 \\
0 & 0 & 0 & 0 & 0 & 0 & 0 \\
0 & 0 & 0 & 0 & 0 & 0 & 1 \\
0 & 0 & 0 & 0 & 0 & -1 & 0 \\
0 & 0 & 0 & 0 & 1 & 0 & 0 \\
0 & 0 & 0 & -1 & 0 & 0 & 0 
\end{array}
\right)
\qquad
C_{10} =\frac 1{\sqrt 3} \left(
\begin{array}{ccccccc}
0 & 0 & -2 & 0 & 0 & 0 & 0 \\
0 & 0 & 0 & 0 & 0 & 0 & 0 \\
2 & 0 & 0 & 0 & 0 & 0 & 0 \\
0 & 0 & 0 & 0 & 0 & -1 & 0 \\
0 & 0 & 0 & 0 & 0 & 0 & -1 \\
0 & 0 & 0 & 1 & 0 & 0 & 0 \\
0 & 0 & 0 & 0 & 1 & 0 & 0 
\end{array}
\right)
$$
$$
 C_{11} = \frac 1{\sqrt 3} \left(
\begin{array}{ccccccc}
0 & 0 & 0 & -2 & 0 & 0 & 0 \\
0 & 0 & 0 & 0 & 0 & 0 & -1 \\
0 & 0 & 0 & 0 & 0 & 1 & 0 \\
2 & 0 & 0 & 0 & 0 & 0 & 0 \\
0 & 0 & 0 & 0 & 0 & 0 & 0 \\
0 & 0 & -1 & 0 & 0 & 0 & 0 \\
0 & 1 & 0 & 0 & 0 & 0 & 0 
\end{array}
\right)
\qquad
C_{12} = \frac 1{\sqrt 3} \left(
\begin{array}{ccccccc}
0 & 0 & 0 & 0 & -2 & 0 & 0 \\
0 & 0 & 0 & 0 & 0 & 1 & 0 \\
0 & 0 & 0 & 0 & 0 & 0 & 1 \\
0 & 0 & 0 & 0 & 0 & 0 & 0 \\
2 & 0 & 0 & 0 & 0 & 0 & 0 \\
0 & -1 & 0 & 0 & 0 & 0 & 0 \\
0 & 0 & -1 & 0 & 0 & 0 & 0 
\end{array}
\right)
$$
$$
C_{13} = \frac 1{\sqrt 3} \left(
\begin{array}{ccccccc}
0 & 0 & 0 & 0 & 0 & -2 & 0 \\
0 & 0 & 0 & 0 & -1 & 0 & 0 \\
0 & 0 & 0 & -1 & 0 & 0 & 0 \\
0 & 0 & 1 & 0 & 0 & 0 & 0 \\
0 & 1 & 0 & 0 & 0 & 0 & 0 \\
2 & 0 & 0 & 0 & 0 & 0 & 0 \\
0 & 0 & 0 & 0 & 0 & 0 & 0 
\end{array}
\right)
\qquad
C_{14} = \frac 1{\sqrt 3} \left(
\begin{array}{ccccccc}
0 & 0 & 0 & 0 & 0 & 0 & -2 \\
0 & 0 & 0 & 1 & 0 & 0 & 0 \\
0 & 0 & 0 & 0 & -1 & 0 & 0 \\
0 & -1 & 0 & 0 & 0 & 0 & 0 \\
0 & 0 & 1 & 0 & 0 & 0 & 0 \\
0 & 0 & 0 & 0 & 0 & 0 & 0 \\
2 & 0 & 0 & 0 & 0 & 0 & 0 
\end{array}
\right)
$$
\normalsize
This basis satisfy the commutation rules summarized in the antisymmetric matrix
$B_{IJ} =[C_I , C_J]$ given in the appendix \ref{Bmatr}.\\[0.5cm]
\indent Among that matrices we can recognize the Lie algebras corresponding 
to the subgroups of $G_2$ mentioned above. 
The first eight matrices generate $Lie(P)$ and they are reminiscent of the 
Gell-Mann matrices.
Moreover, the matrices $\{ C_1 \ , C_2 \ , C_3 \}$ generate $Lie(S)$ and finally 
$\{ C_8 \ , C_9 \ , C_{10} \}$ generate $Lie(\Sigma)$.
 
Since the elements $C_5$ and $C_{11}$ commute, they generate a Cartan 
subalgebra of $Lie(G_2)$ which is the Lie algebra of the maximal torus $T$ of $G_2$.
Notice that the commutators among the basis of $Lie(S)$, $Lie{\Sigma}$ and $Lie(T)$ 
generate the whole basis of $Lie(G_2)$.\\
Our previous observations lead us to the conjecture that a good 
coordinatization for the generic element $g\in G_2$ can be defined by
\eqn
g=\sigma(a_1 , a_2 , a_3) s(a_4 , a_5 , a_6) e^{\sqrt 3 a_7 C_{11}} 
e^{a_8 C_5} u(a_9 , a_{10} , a_{11} ; a_{12} , a_{13} , a_{14}),
\label{coordinatization}
\feqn
where
\eqn
&& s (x , y , z) = e^{x C_3} e^{y C_2} e^{z C_3} \ , \\
&& \sigma (x' , y' , z') = e^{\sqrt 3 x' C_8} e^{\sqrt 3 y' C_9} 
e^{\sqrt 3 z' C_8} \ , \\
&& u(x , y , z ; x' , y' , z') =s (x , y , z) \sigma (x' , y' , z') 
\label{param}
\feqn
are elements respectively of $S$, $\Sigma$ and $SO(4)$.\\
In this paper we prove that this is in fact a good coordinatization for 
$G_2$, which could be used to determine a simple form for the Haar
measure on the group.  
In order to achieve this we will first determine the corresponding invariant 
metric and then compute the range of the coordinates $a_1, a_2, \ldots , 
a_{14}$.


\section{The invariant metric for \boldmath{$G_2$}}
\setcounter{equation}{0}

In order to compute the invariant metric over the Lie group $G_2$ 
we will first show how our coordinatization (\ref{coordinatization}) is
related to the fibration described in the previous section.

It is well known that for a simple group, and therefore in particular for 
$G_2$, the invariant metric is uniquely defined (up to a normalization 
constant) by the Killing form over the algebra. More precisely, the Killing 
form defines a metric (and a Lebesgue measure) over the tangent space to the 
identity, which can be pulled back via the left (or right) multiplication.
If ${f_{IJ}}^K$ are the structure constants of the 
algebra, then the Killing metric has components
\eqn
K_{IJ}=(C_I,C_J) =-k{f_{IL}}^M {f_{JM}}^L \ ,
\feqn
where $k$ is a normalization constant. In our case we find 
$K_{IJ} =16 k \delta_{IJ}$, which suggests to choose $k=\frac 1{16}$ 
conveniently in such a way that the generators 
$\{ C_I \}_{I=1}^{14}$ are orthonormal: $(C_I , C_J ) =\delta_{IJ}$.\\
By right multiplication we can associate to every matrix $C_I$ a vector field
in the tangent bundle. Its dual (defined using the pullback
of the Killing form) is a left-invariant 1-form and the collection of those 
1-forms provides a trivialization of the cotangent bundle. In a coordinate patch 
$\{ w^I \}$ the canonical 1-form J  becomes
\eqn
J = J^I C_I = g^{-1} \frac {\partial g}{\partial w^J} dw^J \ , 
\feqn
so that the invariant metric is
\eqn
ds^2 =g_{IJ} dw^I \otimes dw^J = J^L \otimes J^M (C_L , C_M) \ . 
\feqn 
Therefore, if we define the matrix $\underline J = \{ {J^I}_K \}$ with
components $J^I ={J^I}_K dw^K$ and remember our
normalization, we find for the components of the metric
\eqn
g_{IJ} =\delta_{LM} {J^L}_I {J^M}_J \ ,
\feqn 
This means that the right currents define a $14$--bein over 
the Lie group $G_2$. 
In particular the invariant volume form is 
\eqn
\omega=J^1 \wedge \ldots \wedge J^{14}= \det (\underline J) dx^1 \wedge \ldots \wedge dx^{14}
\label{volform}
\feqn
and the associated Haar measure is 
\eqn
d\mu =\det (\underline J) \prod_{I=1}^{14} dw^I \ .
\label{measure}
\feqn
Now we use our coordinatization (\ref{coordinatization}), which we 
rewrite in the form 
\eqn
g= h(a_1 , a_2 , a_3 , a_4 , a_5 , a_6 , a_7 , a_8 ) u(a_9 , a_{10} , a_{11} 
; a_{12} , a_{13} , a_{14}) \ .
\feqn
Note that $u$ is the generic element of the subgroup $SO(4)$, while $h$ is an element of 
 $S \Sigma T$ 
which is not a subgroup of $G_2$. 
We can express the currents associated to the elements $u$ and $h$ 
respectively as
\eqn
J_u = du u^{-1} =\sum_{i\in A} J_u^i C_i \ , \qquad J_h = h^{-1}dh 
=\sum_{i=1}^{14} J_h^i C_i \ ,
\feqn
where $A=\{ 1,2,3,8,9,10 \}$. Note that for $J_U$ we have chosen
the left currents.
This is because from the orthonormality condition it follows that
\eqn
ds^2 = \sum_{i\in A} \left( J_u^i +J_h^i \right)^2 +\sum_{i\notin A} 
\left( J_h^i \right)^2 \ . \label{decompmetr}
\feqn
This particular form of the metric stress the ralation with the fibration of $G_2$ over 
the variety $\mathcal H$. In effect it shows explicitly the separation between the base and the fiber and 
if we fix the coordinates $(a_1 , a_2 , a_3 , a_4 , a_5 , a_6 , a_7 , a_8 )$,
it reduces to the metric on the fiber $SO(4)$.
On the other hand the term $\sum_{i\notin A} \left( J_h^i \right)^2$ corresponds to the 
acht-bein $\tilde J_h$ obtained after projecting the currents
$J_h$ orthogonally to the fiber, and therefore it has to coincide with 
the metric on the base.\\
This decomposition greatly simplifies the explicit computation of the metric, 
the main task being the computation of $J_h$.\\ 
In order to determine $J_u$ remember that 
$u(x , y , z ; x' , y' , z') =s (x , y , z) \sigma (x' , y' , z')$ 
and that $s$ and $\sigma$ commute, so that in terms of 
$J_s =ds s^{-1}$ and $J_\sigma =d\sigma \sigma^{-1}$, 
we get $J_u =J_s +J_\sigma$ with
\eqn
&& J_S (x,y,z) = [-\sin (2x) dy +\cos (2x) \sin (2y) dz ]C_1 
+[\cos (2x) dy +\sin (2x) \sin (2y) dz] C_2 \cr
&& \qquad \qquad \qquad
+[dx +\cos (2y) dz ]C_3 \ , \cr
&& J_\Sigma (x,y,z) =\sqrt 3 [dx +\cos (2y) dz ]C_8 
+\sqrt 3 [\cos (2x) dy +\sin (2x) \sin (2y) dz] C_9 \cr
&& \qquad \qquad \qquad
+\sqrt 3 [\sin (2x) dy -\cos (2x) \sin (2y) dz ]C_{10} \ .
\feqn

Using Mathematica we have found for the currents $J_h$ the expressions 
written in the Appendix \ref{appj}.\\[0.5cm]
\indent We are now able to compute the Haar measure in our coordinates. In fact, 
if $\underline {\tilde J_h}$ is the $8\times 8$ matrix given
by the acht-bein $\tilde J_h$ and $\underline {J_s} \ ,\underline {J_\sigma}$ 
are the $3\times 3$ matrices associated to the drei-bein $J_s$ and $J_\sigma$
respectively, then 
$\det (\underline J ) =\det (\underline {\tilde J_h}) 
\det (\underline {J_s}) \det (\underline {J_\sigma})$
so that
\eqn
d\mu=27 \sqrt 3 f(2 a_7 \ , 2a_8 ) \sin (2a_2 ) 
\sin (2a_5) \sin (2a_{10}) \sin (2a_{13} )
\prod_{i=1}^{14} da_i \ , \label{invmeasure}
\feqn 
where  
\eqn
f(\alpha, \beta ) &=&\sin (\frac {\beta-\alpha}2 ) \sin (\frac {\beta+\alpha}2 ) 
\sin (\frac {\beta-3\alpha}2 ) \sin (\frac {\beta+3\alpha}2 ) \sin (\alpha) \sin (\beta) \cr
&=&\frac{1}{4} (\cos(\alpha)+\cos(\beta))(\cos(3\alpha)+\cos(\beta))\sin (\alpha) \sin (\beta)
\label{ffunct}
\feqn
This, however, is not the end of the story. We need in fact to determine the
range of the coordiantes which covers the whole group $G_2$, apart from a 
subset of zero measure. To this end we will use a topological argument.


\section{The range of the coordinates}
\setcounter{equation}{0}

Before entering into more details, let us explain our strategy.\\
Looking at the measure (\ref{invmeasure}) one immediately sees that for some 
values of the coordiantes it vanishes. 
This happens for certain values of the angles
$a_2 , a_5 , a_7 , a_8 , a_{10} , a_{13}$. 
Let us suppose we choose the range for these coordinates in such a way that 
it delimits a region where $d \mu$ is vanishing only on the 
boundary. For the other coordinates the range is fixed
 in such a way that each of them goes around a closed 
orbit exactly once. \\
To this end we describe a $14$-dimensional closed cycle $V$ which represents 
an element of the homology group 
$[V] \in H_{14} (G_2, \mathbb{Z})$ \footnote{However for
our purposes it would be enough to consider the homology with rational or 
real coefficients.}.
Using the pairing $B: H^k \times H_k \longrightarrow \mathbb{R}$ given by
\eqn
B( [\xi], [W]) =\int_W \xi \ ,
\feqn
we define the normalized form
\eqn
\tau =B([\omega], [G_2])^{-1} \omega .
\feqn
It is clear that the function $B([\tau], \cdot )$ takes only integer values.
In particular $B([\tau], [W])$  counts the number of times
the cycle $W$ wraps $G_2$.
Once this is known all we need to do is to find how 
to restrict the range of coordinates until we obtain 
a cycle $V$ which wraps $G_2$ once. \\
We are now going to enter into more details and compute $\tau$ in three steps.

\subsection{The evaluation of \boldmath{$B([\omega] , [G_2])$}}

There is a simple way to compute the total volume of a connected simple Lie
group, described by Macdonald in \cite{Burgy}. It works as follows.\\
If $G$ is the group, $t \subset Lie(G)$ a Cartan subalgebra,
and $t_{\mathbb{Z}}$ the integer lattice generated in $t$ by the simple roots, 
then $T =t/t_{\mathbb{Z}}$ is a torus with the same
dimension as $t$. Let $\alpha >0$ denote the positive roots and
$|\alpha|$ their length. \\
Then from Hopf theorem, the rational homology of $G$ is equal to the rational homology of a product 
of odd-dimensional spheres:
$H_* (G,\mathbb{Q}) \sim H_* \left(\prod_{i=1}^k (S^{2i+1})^{n_i},\mathbb{Q} \right) $, where $n_i$ is the 
number of
times the given sphere appears. Let  $Vol(S^{2i+1}) =2\pi^{i+1}/i!$ be 
the volume of the $(2i+1)$-dimensional unit sphere and $Vol({T})$ be the volume of the 
torus computed using the measure induced by the Lebesgue measure 
on the algebra.\\
Then the whole volume determined via the pullback of the Lebesgue measure on
the algebra is
\eqn
Vol(G) =Vol({T}) \cdot \prod_{i=1}^k Vol(S^{2i+1})^{n_i} \cdot
\prod_{\alpha >0} \frac 4{|\alpha|^2} \ .
\feqn
The roots of $G_2$ computed with our choice for the algebra and the 
normalization, are shown in the Appendix \ref{approot}. From the figure we see that 
there are three positive roots of length $2$ and three of length $\frac 2{\sqrt 3}$. 
The torus associated to the simple roots $\alpha_1$ and $\alpha_2$ is 
generated by the coroots $H_1$ and $H_2$. Remembering the relations 
$|H| =2/|\alpha|$, we find $Vol({T})= \frac {\sqrt 3}2$. 
Since $S^3$ and $S^{11}$ are the odd spheres which generate the 
rational homology of $G_2$, we obtain the desired result
\eqn
B(\mu , G_2) =Vol (G_2) =9\sqrt 3 \frac {\pi^8}{20} \ .
\label{Bvol}
\feqn
We are now ready for the next step.

\subsection{The construction of the cycle \boldmath{$V$}}

Let us look at (\ref{coordinatization}): to determine a closed cycle we first
observe that for each one of the two $SU(2)$ subgroups $S$ and $\Sigma$ 
it is possible to choose the range of the coordinates in such a way as to 
cover the whole $3-$sphere. The method to do this is well-known (Euler angles) 
and here we give only the final result:
\eqn
&& 0 \leq a_1 \leq 2 \pi \ , \qquad  0 \leq a_2 \leq  
\frac \pi2 \ ,
\qquad 0 \leq a_3 \leq \pi \ , \cr
&& 0 \leq a_4 \leq 2 \pi \ , \qquad  0 \leq a_5 \leq \frac \pi2 \ ,
\qquad 0 \leq a_6 \leq \pi \ , \cr
&& 0 \leq a_9 \leq 2 \pi \ , \qquad  0 \leq a_{10} \leq \frac \pi2 \ ,
\qquad 0 \leq a_{11} \leq \pi \ , \cr
&& 0 \leq a_{12} \leq 2 \pi \ , \qquad  0 \leq a_{13} \leq \frac \pi2 \ ,
\qquad 0 \leq a_{14} \leq \pi \ . \label{vrange}
\feqn
To complete the cycle we need to determine the range for $a_7$ and $a_8$ in
such a way that $d\mu$ does not vanish. To this end we solve the inequality $f(x,y) >0$ 
and obtain a tiling of the fundamental region
\eqn
(2a_7 \ , 2a_8) \in [0, 2\pi] \times [0, 2\pi] \ ,
\feqn
as we show in the Appendix \ref{apptil}. There, we also prove that every region of the 
tiling gives the same (absolute value) contribution to the measure.

The cycle $V$ is then obtained by choosing any one of these regions, for 
example the one denoted by $B$ in the figure.

\subsection{The evaluation of \boldmath{$\tau$} and the range of the Euler angles for 
\boldmath{$G_2$}}

We can now evaluate the degree of the map $V\rightarrow G_2$. 
Using (\ref{vrange}) for the range of the coordinates, 
(\ref{invmeasure}) for $d\mu$ and (\ref{Bvol}) for $Vol(G_2)$, we easily find 
$B(\tau, V) =16$.
Therefore, our next task is to understand the origin of this factor.

A factor $4$ can be easily accounted for in the following way. 
We have built the cycle $V$ starting from the closed submanifolds 
$S$ and $\Sigma$ corresponding to the two $SU(2)$ embeddings.
Thus, naively, we would expect to find a $SU(2) \times SU(2)$ submanifold
embedded in $G_2$. But a direct inspection shows that this is not exactly true.
In fact, varying for example $a_1\in [0,2 \pi]$ provides a double covering 
of the six-dimensional submanifold obtained by taking $a_1\in [0, \pi]$ and 
$a_2 , \ldots , a_6 $ as in (\ref{vrange}).
This is because the image of $SU(2) \times SU(2)$ in $G_2$ is 
$SO(4)=(SU(2)\times SU(2))/\mathbb{Z}_2$, as previously remarked in 
Section $2$. 

Similarly, we must reduce the range of $a_{12}$ to  $a_{12} \in [0, \pi]$. 
The new cycle we obtain in this way wraps $G_2$ four times.

Now, let us consider the torus 
${T} (a_7 , a_8) :=e^{\sqrt 3 a_7 C_{11}} e^{a_8 C_5}$. 
We need to determine the subgroup of $7\times 7$ orthogonal matrices $A$  
of $SO(4)$, which leaves each element 
${T} (a_7 , a_8)$ invariant under the adjoint action
\eqn
A {T} (a_7 , a_8) A^t ={T} (a_7 , a_8) \ .
\feqn
It turns out that it is a finite group generated by the idempotent 
matrices $\sigma$ ($\sigma=\sigma^{-1}$) and $\eta$ ($\eta=\eta^{-1}$)
\eqn
\begin{array}{rcl}
\sigma=\left(
\begin{array}{ccccccc}
1 & 0 & 0 & 0 & 0 & 0 & 0 \\
0 & -1 & 0 & 0 & 0 & 0 & 0 \\
0 & 0 & -1 & 0 & 0 & 0 & 0 \\
0 & 0 & 0 & 1 & 0 & 0 & 0 \\
0 & 0 & 0 & 0 & 1 & 0 & 0 \\
0 & 0 & 0 & 0 & 0 & -1 & 0 \\
0 & 0 & 0 & 0 & 0 & 0 & -1 
\end{array}
\right)
\qquad \eta =\left(
\begin{array}{ccccccc}
-1 & 0 & 0 & 0 & 0 & 0 & 0 \\
0 & 0 & 1 & 0 & 0 & 0 & 0 \\
0 & 1 & 0 & 0 & 0 & 0 & 0 \\
0 & 0 & 0 & -1 & 0 & 0 & 0 \\
0 & 0 & 0 & 0 & 1 & 0 & 0 \\
0 & 0 & 0 & 0 & 0 & 0 & -1 \\
0 & 0 & 0 & 0 & 0 & -1 & 0 
\end{array}
\right) \ .              
\end{array}
\feqn
Considering the action of $\sigma$ one finds
\eqn
&& g= U(a_4 , a_5 , a_6 ; a_1 , a_2 , a_3 ) {T} (a_7 , a_8)
U(a_9 , a_{10} , a_{11} ; a_{12} , a_{13} , a_{14} ) \cr
&& \ =
U(a_4 , a_5 , a_6 ; a_1 , a_2 , a_3 ) \sigma {T} (a_7 , a_8) \sigma
U(a_9 , a_{10} , a_{11} ; a_{12} , a_{13} , a_{14} ) \cr
&& \ = U(a_4 , a_5 , a_6 +\frac \pi2 ; a_1 , a_2 , a_3 +\frac \pi2 )
{T} (a_7 , a_8)
U(a_9 +\frac \pi2 , a_{10} , a_{11} ; a_{12} +\frac \pi2 , a_{13} , a_{14} ) .
\feqn
Therefore, we can restrict $0\leq a_3 < \frac \pi2$.

Analogously, let us now look at the symmetry generated by $\eta$:
\eqn
&& g= U(a_4 , a_5 , a_6 ; a_1 , a_2 , a_3 ) {T} (a_7 , a_8)
U(a_9 , a_{10} , a_{11} ; a_{12} , a_{13} , a_{14} ) \cr
&& \ =
U(a_4 , a_5 , a_6 ; a_1 , a_2 , a_3 ) \eta {T} (a_7 , a_8) \eta
U(a_9 , a_{10} , a_{11} ; a_{12} , a_{13} , a_{14} ) ,
\feqn
and restrict our attention to the factor $\eta U$ on the right. 
(Similar relations are true for the factor on the left.) A direct 
computation using the explicit expression of the matrices shows that the 
left action of $\eta$ on $U$ is equivalent to the shift
\eqn
&& a_9 \mapsto \frac \pi4 -a_9 \ , \qquad a_{10} \mapsto a_{10} +\frac \pi2 \ ,
\qquad a_{11} \mapsto a_{11} \ , \cr
&& a_{12} \mapsto -\frac \pi4 -a_{12} \ , \qquad a_{13} \mapsto a_{13} 
+\frac \pi2 \ , \qquad a_{14} \mapsto a_{14} \ .
\label{shift}
\feqn
The analysis is more complicated than for the action of $\sigma$, because now
some of the angles are mapped to values which are outside of the 
range (\ref{vrange}) we have fixed. For example, 
$a_{10} +\frac \pi2 \in [\frac \pi2,\pi]$, when $a_{10}\in [0,\frac \pi2]$.
Therefore, we need to use other equivalence relations to map the angles back
to the original region (\ref{vrange}).

In fact, the following symmetries hold for $S$ and $\Sigma$: 
\eqn
&& S[a_9 , a_{10} , a_{11}] \sim S[a_9 +\frac \pi2 , \pi -a_{10} , 
a_{11} +\frac \pi2] \cr
&& \Sigma[a_{12} , a_{13} , a_{14}] \sim \Sigma[a_{12} +\frac \pi2 , 
\pi -a_{13} , a_{14} +\frac \pi2] \ .
\feqn
These are the known symmetries which have been used to determine 
(\ref{vrange}) in the first place.
Thus the action (\ref{shift}) of $\eta$ is equivalent to
\eqn
&& a_9 \mapsto \frac 34 \pi -a_9 \ , \qquad a_{10} \mapsto  
\frac \pi2 -a_{10} \ ,
\qquad a_{11} \mapsto a_{11} +\frac \pi2 \ , \cr
&& a_{12} \mapsto \frac \pi4 -a_{12} \ , \qquad a_{13} \mapsto 
\frac \pi2 -a_{13} \ ,
\qquad a_{14} \mapsto a_{14} +\frac \pi2 \ .
\feqn
where now $a_{10}$ and $a_{13}$ stay inside the allowed intervals.
The next point to notice is that for the remaining coordinates it is 
possible to use similarity relations which does not involve 
$a_{10}$ and $a_{13}$, e.g.
\eqn
S[a_9 , a_{10} , a_{11}] \Sigma[a_{12} , a_{13} , a_{14}]  \sim
S[a_9 +\pi , a_{10} , a_{11}] \Sigma[a_{12} +\pi , a_{13} , a_{14}]
\feqn
and similar symmetries. Moreover, $\eta$ is a linear transformation, 
so that it is enough to restrict the range of one of the angles.
Therefore, luckily, we actually do not need to know the action of $\eta$ 
on the whole set of coordinates. The solution of our problem simply consists in 
restricting the range of either $a_{10}$ or $a_{13}$ in the $SO(4)$ factor 
$U(a_9 , a_{10} , a_{11} ; a_{12} , a_{13} , a_{14} )$ on the right. 
Alternatively, the same result can be achieved by restricting the range 
of either $a_2$ or $a_5$ in the factor $U(a_4 , a_5 , a_6 ; a_1 , a_2 , a_3 )$
on the left. 
Since we prefer a coordinatization which respects the 
fibration described in Section 2 and we want the angles 
$a_9,\ldots,a_{14}$ to span the whole fiber $U(a_9, \ldots , a_{14})$, 
we choose the second option and restrict the range of $a_5$.

Finally, we can summarize our results for the range of the angles 
describing $G_2$:
\eqn
&& 0 \leq a_1 \leq \pi \ , \qquad  0 \leq a_2 \leq  
\frac \pi2 \ ,
\qquad 0 \leq a_3 \leq \frac \pi2 \ , \cr
&& 0 \leq a_4 \leq 2 \pi \ , \qquad  0 \leq a_5 \leq \frac \pi4 \ ,
\qquad 0 \leq a_6 \leq \pi \ , \cr
&& 0 \leq a_9 \leq 2 \pi \ , \qquad  0 \leq a_{10} \leq \frac \pi2 \ ,
\qquad 0 \leq a_{11} \leq \pi \ , \cr
&& 0 \leq a_{12} \leq  \pi \ , \qquad  0 \leq a_{13} \leq \frac \pi2 \ ,
\qquad 0 \leq a_{14} \leq \pi \ , \cr
&& 0 \leq a_7 \leq \frac \pi6 \ , \qquad  3a_7 \leq a_8 \leq \frac \pi2 \ .
\feqn


\section{Conclusions}
\setcounter{equation}{0}

We have found a coordinatization of the $G_2$ group with a one to one 
correspondence between the range of coordinates and a full measure subset 
of the group. In particular, this has allowed us to obtain a quite simple 
expression for the Haar measure, which should make numerical computations 
involving the geometry of $G_2$ much easier, for example in lattice gauge 
theories or in random matrix models.

However, note that to find an Haar measure on the group we would not 
have actually needed the last step of the work, i.e. the determination of the 
correct range of the coordinates which yields an injective map.
 
In fact, if $d\mu$ indicates the measure over the cycle $V$, it is possible 
to simply get an Haar measure $d\tilde \mu$ over the Lie group  $G_2$ by just taking
\eqn
d\tilde \mu := \frac 1{16} d\mu \ ,
\feqn
$16$ being the degree of the map $V\longrightarrow G_2$.

On the other hand the determination of the correct range of coordinates,
which covers the whole $G_2$ wrapping it exactly once
except for a subset of vanishing measure, provides a new 
result, since it determines also a coordinatization of 
the homogeneous space $\mathcal{H}= G_2/SO(4)$ of quaternionic 
subalgebras of octonions.

Moreover, we have also computed the induced metric of $\mathcal H$
\eqn
ds^2_{\mathcal{H}}=\delta_{ab} \tilde J_h^a \otimes \tilde J_h^b \ ,
\feqn
and shown that it is an Einstein metric.
This result is proven in the Appendix \ref{apph}.

\section*{Acknowledgments}

SC and GO would like to thank G.~Berrino, M.~Bertini, S.~Bertini and 
B.~Van~Geemen for helpful conversations.
BLC would like to thank O.~Ganor, B.~Zumino, M.~Papucci for very
useful discussions. 
ADV would like to thank P.~Cotta~Ramusino for interesting discussions on the 
octonions and exceptional Lie groups.
BLC is supported by the INFN under grant 8930/01.
This work was supported in part by the Director, Office of Science, Office of
High Energy and Nuclear Physics, of the U.S. Department of Energy 
under Contract DE-AC03-76SF00098 and in part by the US Department of Energy, 
Grant No.DE-FG02-03ER41262. 
The work of SC is supported by INFN, COFIN prot. 2003023852\_008 and the 
European Commission RTN program MRTN--CT--2004--005104, in which SC is
associated to the University of Milano--Bicocca.

\newpage
\begin{appendix}

\section{Commutators}
\setcounter{equation}{0}

\label{Bmatr}
\begin{equation*}
\begin{array}{rcl}
B= && \left(
\begin{array}{cccccccc}
0 & 2C_3 & -2C_2 & C_7 & C_6 & -C_5 & -C_4 & 0 \\
* & 0 & 2C_1 & -C_6 & C_7 & C_4 & -C_5 & 0 \\
* & * & 0 & C_5 & -C_4 & C_7 & -C_6 & 0 \\
* & * & * & 0 & C_3 +\sqrt 3 C_8 & -C_2 & C_1 & -\sqrt 3 C_5 \\
* & * & * & * & 0 & C_1 & C_2 & \sqrt 3 C_4 \\
* & * & * & * & * & 0 & C_3 -\sqrt 3 C_8 & \sqrt 3 C_7 \\
* & * & * & * & * & * & 0 & -\sqrt 3 C_6 \\
* & * & * & * & * & * & * & 0 \\
* & * & * & * & * & * & * & * \\
* & * & * & * & * & * & * & * \\
* & * & * & * & * & * & * & * \\
* & * & * & * & * & * & * & * \\
* & * & * & * & * & * & * & * \\
* & * & * & * & * & * & * & * \\
\end{array} \right. \\
&& \left.
\begin{array}{cccccc}
0 & 0 & C_{14} & C_{13} & -C_{12} & -C_{11} \\
0 & 0 & -C_{13} & C_{14} & C_{11} & -C_{12} \\
0 & 0 & C_{12} & -C_{11} & C_{14} & -C_{13} \\
-C_{14} & -C_{13} & 0 & 0 & C_{10} & C_9 \\
C_{13} & -C_{14} & 0 & 0 & -C_9 & C_{10} \\
-C_{12} & C_{11} & -C_{10} & C_9 & 0 & 0 \\
C_{11} & C_{12} & -C_9 & -C_{10} & 0 & 0 \\
\frac 2{\sqrt 3} C_{10} & -\frac 2{\sqrt 3}
C_9 & -\frac 1{\sqrt 3} C_{12} & \frac 1{\sqrt 3} C_{11} & \frac 1{\sqrt 3}
C_{14} & -\frac 1{\sqrt 3} C_{13} \\
0 & \frac 2{\sqrt 3} C_8 & C_7 -\frac 2{\sqrt 3}
C_{14} & \frac 2{\sqrt 3} C_{13} -C_6 & C_5 -\frac 2{\sqrt 3} C_{12} &
\frac 2{\sqrt 3} C_{11} -C_4 \\
* & 0 & \frac 2{\sqrt 3} C_{13} +C_6 &
\frac 2{\sqrt 3} C_{14} +C_7 & -\frac 2{\sqrt 3} C_{11} -C_4 &
-\frac 2{\sqrt 3} C_{12} +C_5 \\
* & * & 0 &
-\frac 1{\sqrt 3} C_{8} +C_3 & \frac 2{\sqrt 3} C_{10} -C_2 &
-\frac 2{\sqrt 3} C_{9} +C_1 \\
* & * & * & 0 & \frac 2{\sqrt 3} C_{9} +C_1 & \frac 2{\sqrt 3} C_{10} +C_2 \\
* & * & * & * & 0 & \frac 1{\sqrt 3} C_{8} +C_3 \\
* & * & * & * & * & 0
\end{array}
\right)
\end{array}
\end{equation*}
\section{The left-invariant 1-forms \boldmath{$J_h$}}
\label{appj}
\setcounter{equation}{0}

{\small
\eqn
&& J_h^1 =\left[ \cos^3 \left( {a_7} \right) \sin (2a_6)\cos a_8 -\sin^3 
\left( {a_7} \right)
\cos (2a_6)\sin a_8 \right] da_5 \cr
&& \ \qquad -\sin (2a_5) 
\left[ \cos^3 \left( {a_7} \right) \cos (2a_6)\cos a_8 +\sin^3 \left( {a_7} 
\right)
\sin (2a_6)\sin a_8 \right] da_4 \cr
&& \ \qquad +\frac {3}2 \sin \left( 2a_7 \right) 
\left[ \cos \left( 2a_3 \right) \sin \left( a_7 \right) \cos a_8 
+\sin \left( 2a_3 \right) \cos \left( a_7 \right) \sin a_8 \right] da_2 \cr
&& \ \qquad +\frac 32 \sin \left( 2a_7 \right) \sin \left( 2a_2 \right)
\left[ \sin \left( 2a_3 \right) \sin \left( a_7 \right) \cos a_8 \right. \cr
&& \ \qquad \left. 
-\cos \left( 2a_3 \right) \cos \left( a_7 \right) \sin a_8 \right] da_1\\
&& J_h^2 =\left[ \cos^3 \left( a_7 \right) \cos (2a_6)\cos a_8 -\sin^3 
\left( a_7 \right)
\sin (2a_6)\sin a_8 \right] da_5 \cr
&& \ \qquad +\sin (2a_5) 
\left[ \cos^3 \left( a_7 \right) \sin (2a_6)\cos a_8 +\sin^3 \left( a_7 \right)
\cos (2a_6)\sin a_8 \right] da_4 \cr
&& \ \qquad -\frac 32 \sin \left( 2a_7 \right) 
\left[ \sin \left( 2a_3 \right) \sin \left( a_7 \right) \cos a_8 
+\cos \left( 2a_3 \right) \cos \left( a_7 \right) \sin a_8 \right] da_2 \cr
&& \ \qquad +\frac 32 \sin \left( 2a_7 \right) \sin \left( 2a_2 \right)
\left[ \cos \left( 2a_3 \right) \sin \left( a_7 \right) \cos a_8 \right. \cr
&& \ \qquad \left. 
-\sin \left( 2a_3 \right) \cos \left( a_7 \right) \sin a_8 \right] da_1\\
&& J_h^3 =\frac 14 \left( 3\cos \left( 2a_7 \right) +\cos (2a_8) \right) 
\left( 
da_6 +cos (2a_5) da_4 \right) \cr
&& \ \qquad -\frac {\sqrt 3}4 \left( \cos \left( 2a_7 \right) -\cos (2a_8) 
\right) \left( 
da_3 +cos \left( 2a_2 \right) da_1 \right) \\
&& J_h^4 =-\frac 12 \sin (2a_8) da_6 -\frac 12 \cos (2a_5) \sin (2a_8 )da_4 
-\frac 32 \sin (2a_8) da_3 \cr
&& \ \qquad -\frac 32
\sin (2a_8 ) \cos \left( 2a_2 \right) da_1 \\
&& J_h^5 =da_8 \\
&& J_h^6 =\left[ \sin^3 \left( a_7 \right) \cos (2a_6)\cos a_8 +\cos^3 
\left( a_7 \right)
\sin (2a_6)\sin a_8 \right] da_5 \cr
&& \ \qquad +\sin (2a_5) 
\left[ \sin^3 \left( a_7 \right) \sin (2a_6)\cos a_8 -\cos^3 \left( a_7 \right)
\cos (2a_6)\sin a_8 \right] da_4 \cr
&& \ \qquad +\frac 32 \sin \left( 2a_7 \right) 
\left[ -\sin \left( 2a_3 \right) \cos \left( a_7 \right) \cos a_8 
+\cos \left( 2a_3 \right) \sin \left( a_7 \right) \sin a_8 \right] da_2 \cr
&& \ \qquad +\frac 32 \sin \left( 2a_7 \right) \sin \left( 2a_2 \right)
\left[ \cos \left( 2a_3 \right) \cos \left( a_7 \right) \cos a_8 \right. \cr
&& \ \qquad \left. 
+\sin \left( 2a_3 \right) \sin \left( a_7 \right) \sin a_8 \right] da_1\\
&& J_h^7 =\left[ \sin^3 \left( a_7 \right) \sin (2a_6)\cos a_8 +\cos^3 
\left( a_7 \right)
\cos (2a_6)\sin a_8 \right] da_5 \cr
&& \ \qquad -\sin (2a_5) 
\left[ \sin^3 \left( a_7 \right) \cos (2a_6)\cos a_8 -\cos^3 \left( a_7 \right)
\sin (2a_6)\sin a_8 \right] da_4 \cr
&& \ \qquad +\frac 32 \sin \left( 2a_7 \right) 
\left[ \cos \left( 2a_3 \right) \cos \left( a_7 \right) \cos a_8 
-\sin \left( 2a_3 \right) \sin \left( a_7 \right) \sin a_8 \right] da_2 \cr
&& \ \qquad +\frac 32 \sin \left( 2a_7 \right) \sin \left( 2a_2 \right)
\left[ \sin \left( 2a_3 \right) \cos \left( a_7 \right) \cos a_8 \right. \cr
&& \ \qquad \left. 
+\cos \left( 2a_3 \right) \sin \left( a_7 \right) \sin a_8 \right] da_1 \\
&& J_h^{11} =da_7 \\
&& J_h^{12} =\frac {\sqrt 3}2 \sin \left( 2a_7 \right) da_6 +\frac {\sqrt 3}2 
\cos (2a_5) \sin \left( 2a_7 \right) da_4 -\frac {\sqrt 3}2 \sin 
\left( 2a_7 \right) da_3 \cr
&& \ \qquad -\frac {\sqrt 3}2
\sin \left( 2a_7 \right) \cos \left( 2a_2 \right) da_1 \\
&& J_h^{13} =-\frac {\sqrt 3}2 \sin \left( 2a_7 \right)
\left[ \cos \left( a_7 \right) \cos (2a_6) \cos a_8 +\sin \left( a_7 \right)
\sin (2a_6) \sin a_8 \right] da_5 \cr
&& \ \qquad
+\frac {\sqrt 3}2 \sin \left( 2a_7 \right) \sin ( 2a_5)
\left[ -\cos \left( a_7 \right) \sin (2a_6) \cos a_8 +\sin \left( a_7 \right)
\cos (2a_6) \sin a_8 \right] da_4 \cr
&& \ \qquad +\sqrt 3 \left[ \sin \left( 2a_3 \right) \sin \left( a_7 \right) 
\left( 3\sin^2 \left( a_7 \right)-2 \right) \cos a_8 \right. \cr
&& \ \qquad \left.
- \cos \left( 2a_3 \right) \cos \left( a_7 \right) 
\left( 3\cos^2 \left( a_7 \right)-2 \right) \sin a_8 \right] da_2 \cr
&& \ \qquad -\sqrt 3 \sin \left( 2a_2 \right) \left[ \cos \left( 2a_3 \right) 
\sin \left( a_7 \right) 
\left( 3\sin^2 \left( a_7 \right)-2 \right) \cos a_8 \right. \cr
&& \ \qquad \left.
+ \sin \left( 2a_3 \right) \cos \left( a_7 \right) 
\left( 3\cos^2 \left( a_7 \right)-2 \right) \sin a_8 \right] da_1 \\
&& J_h^{14} =\frac {\sqrt 3}2 \sin \left( 2a_7 \right)
\left[ \cos \left( a_7 \right) \sin (2a_6) \cos a_8 +\sin \left( a_7 \right)
\cos (2a_6) \sin a_8 \right] da_5 \cr
&& \ \qquad
-\frac {\sqrt 3}2 \sin \left( 2a_7 \right) \sin ( 2a_5)
\left[ \cos \left( a_7 \right) \cos (2a_6) \cos a_8 -\sin \left( a_7 \right)
\sin (2a_6) \sin a_8 \right] da_4 \cr
&& \ \qquad +\sqrt 3 \left[ \cos \left( 2a_3 \right) \sin \left( a_7 \right) 
\left( 3\sin^2 \left( a_7 \right)-2 \right) \cos a_8 \right. \cr
&& \ \qquad \left.
- \sin \left( 2a_3 \right) \cos \left( a_7 \right) 
\left( 3\cos^2 \left( a_7 \right)-2 \right) \sin a_8 \right] da_2 \cr
&& \ \qquad +\sqrt 3 \sin \left( 2a_2 \right) \left[ \sin \left( 2a_3 \right) 
\sin \left( a_7 \right) 
\left( 3\sin^2 \left( a_7 \right)-2 \right) \cos a_8 \right. \cr
&& \ \qquad \left.
+ \cos \left( 2a_3 \right) \cos \left( a_7 \right) 
\left( 3\cos^2 \left( a_7 \right)-2 \right) \sin a_8 \right] da_1
\feqn
}

\section{The root system}
\setcounter{equation}{0}
\label{approot}
Here we show the roots computed using $C_5$ and $C_{11}$ normalized to $1$. 
The long roots have length $2$ and the short ones have length $2/\sqrt 3$.\\
\setlength{\unitlength}{1mm}
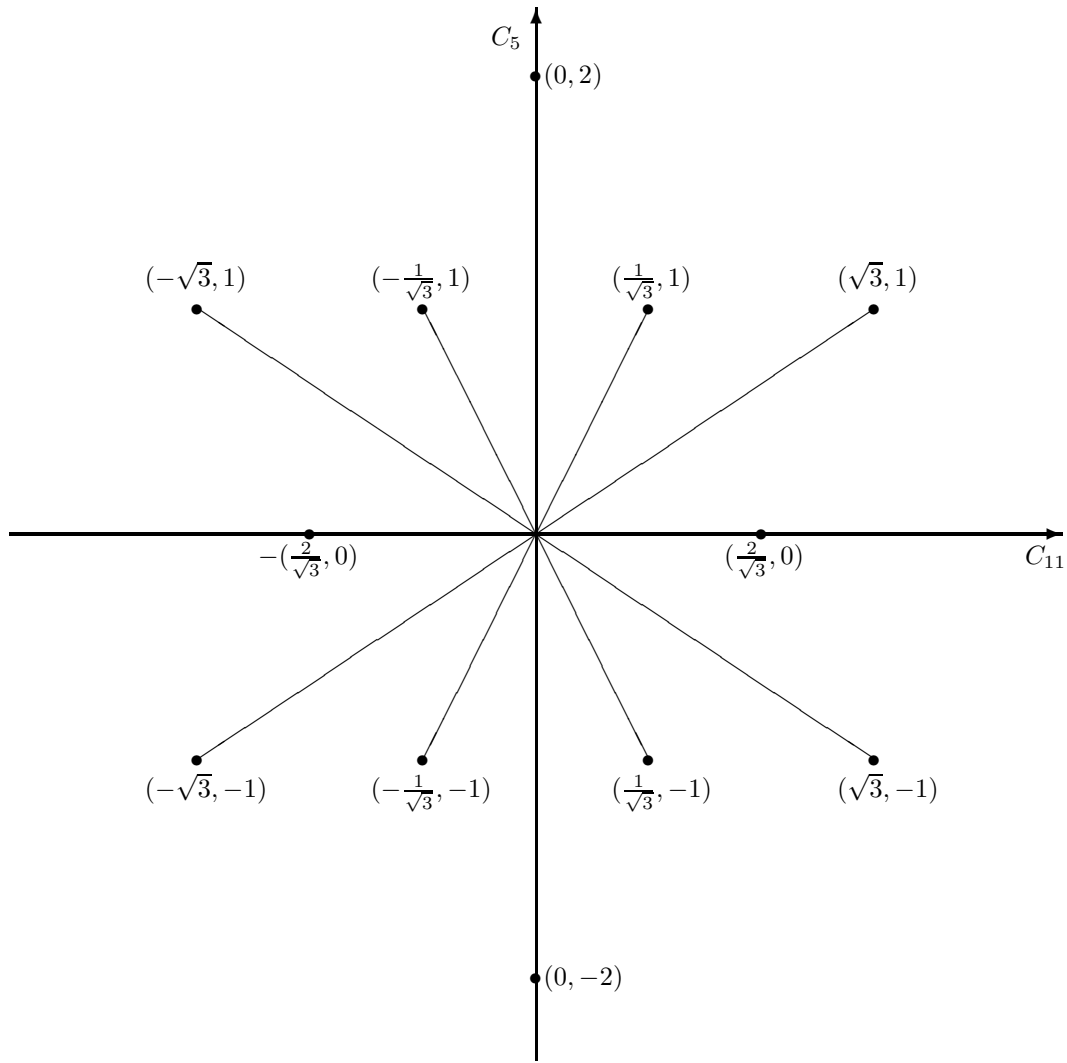
\begin{picture}(140,150)
\put(69,130){$\bullet$}
\put(69,10){$\bullet$}
\put(99,69){$\bullet$}
\put(39,69){$\bullet$}
\put(84,99){$\bullet$}
\put(54,99){$\bullet$}
\put(114,99){$\bullet$}
\put(24,99){$\bullet$}
\put(84,39){$\bullet$}
\put(54,39){$\bullet$}
\put(114,39){$\bullet$}
\put(24,39){$\bullet$}

\put(64,135){$C_5$}
\put(135,66){$C_{11}$}
\put(71,130){$(0,2)$}
\put(71,10){$(0,-2)$}
\put(95,66){$(\frac 2{\sqrt 3},0)$}
\put(33,66){$-(\frac 2{\sqrt 3},0)$}
\put(80,103){$(\frac 1{\sqrt 3},1)$}
\put(48,103){$(-\frac 1{\sqrt 3},1)$}
\put(110,103){$(\sqrt 3 ,1)$}
\put(18,103){$(-\sqrt 3 ,1)$}
\put(80,35){$(\frac 1{\sqrt 3},-1)$}
\put(48,35){$(-\frac 1{\sqrt 3},-1)$}
\put(110,35){$(\sqrt 3 ,-1)$}
\put(18,35){$(-\sqrt 3 ,-1)$}

\thicklines
\put(0,70){\vector(1,0){140}}
\put(70,0){\vector(0,1){140}}
\thinlines
\put(70,70){\line(1,2){15}}
\put(70,70){\line(-1,-2){15}}
\put(70,70){\line(-1,2){15}}
\put(70,70){\line(1,-2){15}}
\put(70,70){\line(3,2){45}}
\put(70,70){\line(-3,2){45}}
\put(70,70){\line(3,-2){45}}
\put(70,70){\line(-3,-2){45}}
\end{picture}

\newpage
\section{The range for \boldmath{$a_7$} and \boldmath{$a_8$}}
\setcounter{equation}{0}
\label{apptil}
Here we show a plot of the fundamental region for the variables $a_7$ and $a_8$, 
which is determined by the condition $f(\alpha,\beta)>0$ where $f$ is given in (\ref{ffunct}). 
We obtain a tiling
of the torus in $24$ triangles, over which the sign of the measure 
alternates, starting with a positive sign in the region $B$. In the edges the measure
vanishes.\\
\setlength{\unitlength}{1mm}
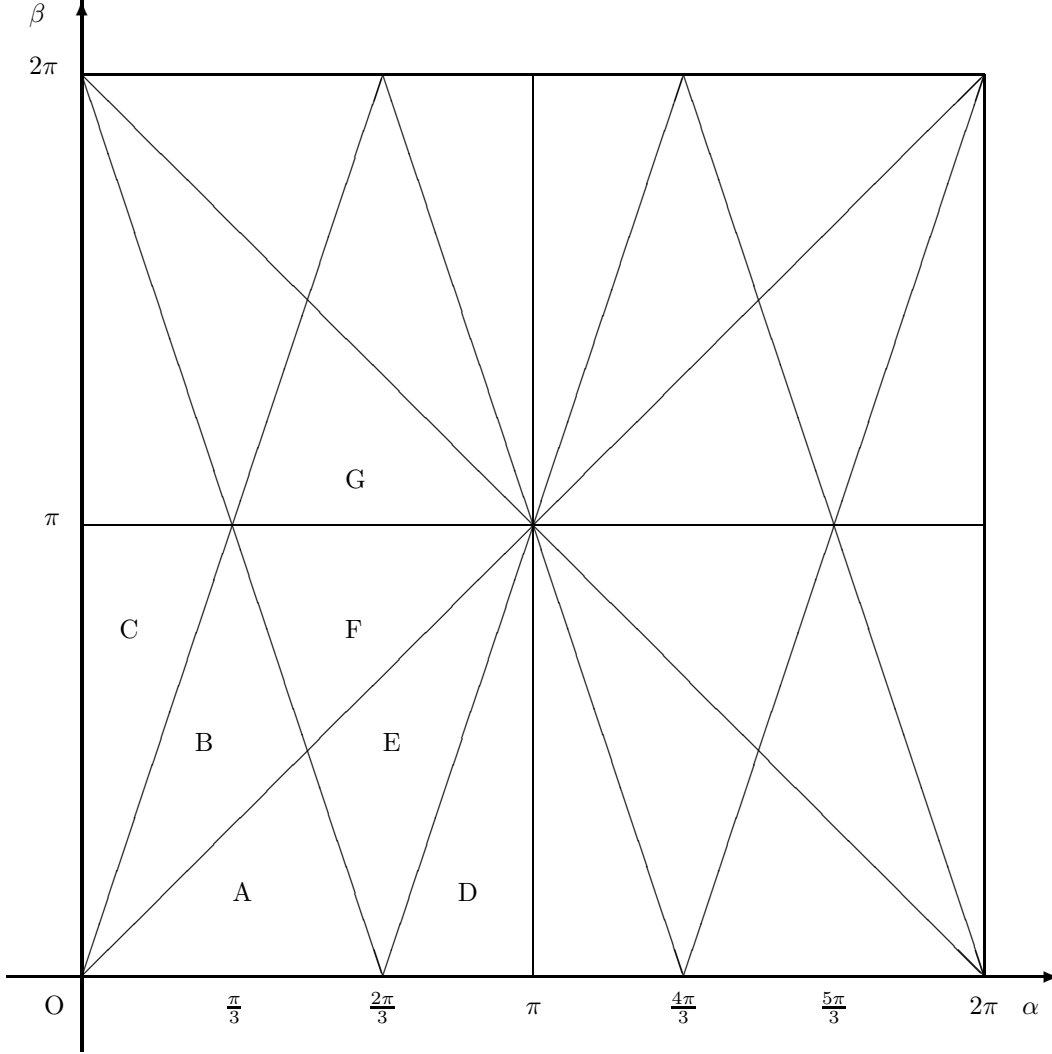
\begin{picture}(140,140)
\put(3,137){$\beta$}
\put(135,5){$\alpha$}
\put(3,130){$2\pi$}
\put(5,70){$\pi$}
\put(5,5){O}
\put(29,5){$\frac \pi3$}
\put(48,5){$\frac {2\pi}3$}
\put(69,5){$\pi$}
\put(88,5){$\frac {4\pi}3$}
\put(108,5){$\frac {5\pi}3$}
\put(128,5){$2\pi$}
\put(25,40){B}
\put(30,20){A}
\put(15,55){C}
\put(50,40){E}
\put(60,20){D}
\put(45,55){F}
\put(45,75){G}

\thicklines
\put(0,10){\vector(1,0){140}}
\put(10,0){\vector(0,1){140}}
\put(10,130){\line(1,0){120}}
\put(130,10){\line(0,1){120}}
\thinlines
\put(10,70){\line(1,0){120}}
\put(70,10){\line(0,1){120}}
\put(10,10){\line(1,3){40}}
\put(50,10){\line(1,3){40}}
\put(90,10){\line(1,3){40}}
\put(10,10){\line(1,1){120}}
\put(10,130){\line(1,-1){120}}
\put(10,130){\line(1,-3){40}}
\put(50,130){\line(1,-3){40}}
\put(90,130){\line(1,-3){40}}
\end{picture}
\\
We now show that any sector of this tiling gives exactly the same 
contribution (obviously up to a sign) to the volume of $G_2$.
To this end we describe some symmetry property of the function $f(\alpha,\beta)$.  
The translation symmetries $f(\alpha,\beta)=f(\alpha+\pi,\beta+\pi)=f(\alpha-\pi,\beta+\pi)$
and the reflection symmetries $-f(\alpha,\beta)=f(-\alpha,\beta)=f(\alpha,-\beta)$ allows to restrict us 
to the square $[0,\pi]\times [0,\pi]$. Moreover, the translations gives the equivalence of the triangles
$A$, $B$ and $D$ with  $F$, $E$ and $C$ respectively.\\
At this point we are left with only three different kinds of triangles: $A$, $B$ and $D$. 
The symmetry $f(\alpha,\beta)=f(\frac{\alpha+\beta}{2},\frac{3\alpha-\beta}{2})$
maps $A$ to $B$, $B$ to $A$ and  $D$ to $G$.  
This proves that we can choose whatever triangle, for example $A$, as the fundamental region.

\newpage
\section{A metric for \boldmath{$\mathcal {H}$}}
\setcounter{equation}{0}
\label{apph}
Here we give the expression for the metric on $\mathcal {H}$ 
induced by the metric on $G_2$, and show that it is an Einstein metric.
Let us introduce the $1$--forms
\eqn
&& I_1(x,y,z):=\sin(2y) \cos (2z) dx -\sin (2z) dy \ , \cr
&& I_2(x,y,z):=\sin(2y) \sin (2z) dx +\cos (2z) dy \ , \cr
&& I_3(x,y,z):=dz+\cos(2y)dx \ .
\feqn
Thus we can write
{\small
\eqn
&& ds^2_{\mathcal {H}} =
da_8^2 +da_7^2 +\left[ \sin^2 a_8 \cos^2 a_7 +\cos^2 a_8 \sin^2 a_7 \right]
\left( da_5^2 +\sin^2 (2a_5) da_4^2  \right. \cr
&& \qquad \ \left. +3da_2^2 +3\sin^2 2a_2 da_1^2 \right)\cr
&& \qquad \ \frac 12 \cos (2a_8) \cos 2a_7 \sin^2 2a_7
\left\{ \left[ I_1 (a_4 , a_5 , a_6 ) +3 I_2 (a_1 , a_2 ,a_3 ) \right]^2
\right. \cr
&& \qquad \ \left. +\left[ I_2 (a_4 , a_5 , a_6 ) -3 I_1 (a_1 , a_2 ,
a_3 ) \right]^2 \right\} \cr
&& \qquad \ +\frac 34 \sin^2 2a_7
\left[ I_3 (a_4 , a_5 , a_6 ) -I_3 (a_1 , a_2 ,a_3 ) \right]^2 \cr
&& \qquad \ +\frac 14 \sin^2 (2a_8) 
\left[ I_3 (a_4 , a_5 , a_6 ) +3I_3 (a_1 , a_2 ,a_3 ) \right]^2 \ .
\label{hmetric}
\feqn
}
We will now compute the curvature of such a metric. Let us use 
capital indices for the full algebra, $I=1,\ldots \ , 14$,
Latin indices for the $so(4)$ subalgebra, $i\in \{ 1,2,3,8,9,10 \} $, 
and Greek indices for the complementary elements,
$\alpha \in \{4 , 5 , 6 , 7 , 11 , 12 , 13 , 14 \}$. 
Let $f_{IJK}$ be the structure constants with one index lowered
through the identity matrix. It is then clear that the non vanishing 
structure constants are those with either none or two Greek 
indices: 
$f_{ijk}$, $f_{\alpha \beta k}$ and permutations. Using this fact and the 
Maurer-Cartan equations\footnote{We use the following notations:
The spin connection is uniquely defined by 
$d\tilde J^\alpha_h =-{\omega^\alpha}_\beta \tilde J^\beta_h $ and 
$\omega_{\alpha \beta}=-\omega_{\beta\alpha}$; the Riemann tensor 
field is then ${\mbox{\cal {R}}^\alpha}_\beta 
=d{\omega^\alpha}_\beta +{\omega^\alpha}_\gamma \wedge {\omega^\gamma}_\beta$
with components ${{R_{\alpha \beta}}^\gamma}_\delta$ 
such that ${\mbox{\cal {R}}^\gamma}_\delta =\frac 12 
{{R_{\alpha \beta}}^\gamma}_\delta J^\alpha_h \wedge J^\beta_h $}, 
the components of the Riemann tensor with respect to the acht-bein are found
using only the algebra. \\
In fact by construction we have that (\ref{hmetric}) takes the form
$ds^2_{\mathcal{H}} =g_{\alpha \beta } da^\alpha \otimes da^\beta$, with
\eqn
g_{\alpha \beta } =\tilde J_\alpha^\gamma \tilde J_\beta^\delta 
\delta_{\gamma \delta} \ . 
\feqn
Let us note that, using notations as in section 3, we have
\eqn
J_h =\tilde J_h +J_h^i C_i \ . \label{semplifica}
\feqn
On the other hand, using $J_h = h^{-1} dh$ one finds
\eqn
dJ_h =-J_h \wedge J_h =-\frac 12 J^I \wedge J^J {f_{IJ}}^K C_K \ .
\feqn
The spin connection can be determined computing $d\tilde J_h$ from 
(\ref{semplifica}).
Being $\tilde J_h^\alpha =J_h^\alpha$ we can omit the tilde in what follows
and write
\eqn
dJ_h^\alpha =-\frac 12 J^I J^J {f_{IJ}}^\alpha =
-\frac 12 J^\beta J^\gamma {f_{\beta \gamma}}^\alpha -J^i
J^\beta {f_{i\beta}}^\alpha   \ ,
\feqn
where we used the properties of the structure constants. From this we can read
the expression for the spin connection one-form, which can be written in the 
form
\eqn
{\omega^\alpha}_\beta =J_h^I {f_{I\beta}}^\alpha -\frac 12 J_h^\gamma
{f_{\gamma\beta}}^\alpha \ .
\feqn
The curvature tensor can be then computed directly
\eqn
{\mbox{\cal {R}}^\alpha}_\beta = J_h^\lambda J_h^\mu
\left[ -{f_{\lambda i}}^\alpha {f_{\mu \beta}}^i +\frac 14 {f_{\lambda\mu}}^\nu
{f_{\nu \beta}}^\alpha -\frac 14 {f_{\mu \nu}}^\alpha {f_{\lambda \beta}}^\nu
-{f_{\lambda \nu}}^\alpha {f_{\mu\beta}}^\nu  \right] \ ,
\feqn
or in components
\eqn
{{R_{\alpha \beta}}^\gamma}_\delta =-{f_{\alpha I}}^\gamma 
{f_{\beta \delta}}^I +{f_{\beta I}}^\gamma {f_{\alpha \delta}}^I
+\frac 12 {f_{\alpha \beta}}^I {f_{I \delta}}^\gamma -\frac 14 
{f_{\beta \delta}}^I {f_{I \alpha}}^\gamma
+\frac 14 {f_{\alpha \delta}}^I {f_{I \beta}}^\gamma \ .
\feqn
The Ricci tensor $\rho_{\alpha\beta} :={{R_{\gamma \alpha}}^\gamma}_\beta$ 
is then
\eqn
\rho_{\alpha \beta} = \frac 14 {f_\alpha}^{IJ} f_{\beta IJ} +\frac 12 
{f_{\alpha}}^{\gamma i} f_{\beta \gamma i}
+\frac 12 {f_{\beta}}^{\gamma i} f_{\alpha \gamma i} \ .
\feqn
The explicit form of the structure constants in our base then yields
\eqn
\rho_{\alpha \beta} =8\delta_{\alpha \beta} \ ,
\feqn
or in curvilinear coordinates 
$\rho_{\mu \nu} dx^\mu \otimes dx^\nu =8 ds_{\mathcal{H}}^2$.
\end{appendix}

\end{document}